\documentstyle{article}
\begin{document}

\centerline{\Large \bf A CATEGORICAL CONSTRUCTION OF 4D }
\smallskip

\centerline{\Large\bf  TOPOLOGICAL QUANTUM FIELD THEORIES}
\bigskip

\centerline{\large Louis Crane\footnote{Supported by NSF grant
DMS-9106476} and David Yetter}
\bigskip

\centerline{Department of Mathematics}
\centerline{Kansas State University}
\centerline{Manhattan, KS 66506}
\bigskip

{\bf 1. Introduction}
\smallskip

In recent years, it has been discovered that invariants of three
dimensional topological objects, such as links and three dimensional
manifolds, can be constructed from various tools of mathematical
physics, such as Von Neumann algebras [1], Quantum Groups [2], and
Rational Conformal Field Theories [3].

Since these different structures lead to the same 3D invariants,
it is natural to wonder how they are related. A fundamental
connection is that they all give rise to the same special tensor
categories, which act as expressions of ``quantum symmetry''
in the very different physical settings [4]. It is therefore
very important that the 3D invariants can all be reconstructed from
a tensor category with the appropriate properties, called a
``modular tensor category (MTC) [3,5].'' The invariants have the property
that they factorize nicely if the manifold or link is cut along a
surface, which we express by saying that they form a Topological
Quantum Field Theory, or TQFT.

Thus the theorem proven in [3] states that a modular tensor
category gives rise to a 3D TQFT. This represents a remarkable
convergence with [6], in which it was realized that a suitable
categorical structure would give rise to 3D topological information,
although without the examples from physics.

The original suggestion to look for TQFT's appears in (not
mathematically rigorous) work of Atiyah [7] and Witten [8]. Atiyah
devotes more attention to the 4D than the 3D situation, and poses
the question whether the two are related; or more concretely, whether
the invariants of Donaldson and Jones are related.
Witten actually works in the 4D situation, and formally suggests
that Donaldson's invariant can be fitted into a 4D TQFT.

At this point, there is no mathematical construction of the 4D TQFT
envisioned in [7] and [8]. Donaldson-Floer theory has so far eluded
the efforts of the strongest of analysts [9].

It is therefore clear that a 4D construction of a TQFT paralleling
the 3D categorical one would have very great implications for
mathematics. As we shall mention in the conclusion section of this
paper, there may very well be important physical implications also.

The construction we describe in this paper is a considerable step in
this direction. Following a suggestion of Ooguri [10], we offer an
expression which gives an invariant of a 4-manifold. The expression
is reminiscent of the invariant of Viro and Turaev [11], and depends
on a triangulation of the manifold. The relative form for a manifold
with boundary, and hence the TQFT, follow directly.

We believe that the invariants, which are constructed from a
modular tensor category within the 3 skeleton of the 4 manifold,
are not the most general, and that the construction should be
extended to a more general one, in which the MTC is replaced by a
suitable 2-category.

For the purposes of this paper, all manifolds are compact, oriented,
and smooth (=PL for D=4).

\bigskip

{\bf 2. The Construction}

\smallskip

In [10], Ooguri, motivated by some ideas in combinatorial quantum
gravity, proposed a formal expression for an invariant for
4-manifolds.
The expression he proposed was divergent, but he
suggested that replacing the representations of a lie group by
those of a quantum group at a root of unity might regularize it.

Since the representations of a quantum group at a root of unity form
a MTC, it is natural to recast Ooguri's suggestion in the context
of [3]. We discovered that a rather small modification of
Ooguri's expression sufficed to produce an invariant of 4-manifolds.

The expression which we found to produce a topological invariant is
a sum of products of contributions for each simplex of the triangulation.

In order to define the invariant, we need to split each tetrahedron
of the triangulation, so that the faces are grouped into two pairs.
Geometrically, we think of this as cutting the four holed sphere
which results from thickening the 1-skeleton of the dual triangulation
of the tetrahedron into two trinions (three holed spheres). See Figure
1.

\vspace*{2in}

\centerline{Figure 1}
\smallskip

The surfaces which result from joining these 4-holed spheres
together are very  important in our definition. Since they reside in
the 3 skeleton of the triangulation, we can often regard them as
boundaries of 3 manifolds, so that the theorems about invariants of
manifolds with boundary in [3] apply.

We can think of a labelling of the faces of the triangulation as
contributing a labeling to the holes of the 4-holed spheres. In order
to pick a basis for the vector space which the MTC attaches to the
four holed sphere, we must cut it along a circle into trinions, and
sum over all labelings of the internal cut by irreducible objects
of the MTC.( If we change cuts, a matrix called the fusion matrix
relates the two bases we obtain.)

If we glue together all the 4-holed spheres from the tetrahedra
on the boundary of one 4-simplex, we obtain a surface imbedded in
$S^3$. See Figure 2.

\vspace*{3in}
\centerline{Figure 2.}
\smallskip

If we pick a vector in the vector space which is assigned to this
surface by the TQFT associated to the MTC, then, since the exterior of
the surface in $S^3$ is a 3-manifold with boundary the surface, we can
obtain an invariant number. In order to describe a vector in the
space associated to the surface, we need to cut the surface into
trinions and label each cut with an irreducible object.
In a general MTC, we need to pick vectors in the space of intertwining
operators as well.

Such a decomposition requires 15 cuts, 10 at faces and 5 inside
tetrahedra. We call the invariant we obtain a generalized 15J symbol.
Such an invariant can also be written by combining generalized
Clebsch-Gordon coefficients, analogously to the classical 15J symbols,
except that care must be taken to describe the embedding of the
surface (thought of as a thickened graph) correctly, by using the
braiding of the category.

Thus, for each 4-simplex, we have a set of numbers, called generalized
15J symbols, corresponding to labellings of the faces and tetrahedra
of the triangulation with irreducibles in the category, and a choice
of two intertwining operators for each tetrahedron.

Our recipe for an invariant is as follows: for each labelling of the
faces of the triangulation by an irreducible object of the
category, each labelling of the cut in each tetrahedron by an
irreducible,
and each labelling of all the resulting trinions by intertwining
operators, we take the product of all the 15J symbols
of the 4-simplices of the triangulation, and correction factors
corresponding to lower dimensional simplices. We then sum over
all labelling.

The result is a topological invariant of the 4 manifold.

This expression is very similar to Ooguri's, except that the
generalization of the 15J symbols to a MTC requires topological
data, because of the braiding in the category. (As we explain below, our
more technical definition of the invariant, which uses an ordering of
the vertices of the triangulation, makes Ooguri's 6J symbols trivial.)

\bigskip

\centerline{\bf 3. Proof of Topological Invariance}

\smallskip

In the special case of quantum SU(2), there are (projectively) unique
intertwiners,
and the various symbols admit explicit expressions in closed form [12].
In the interest of clarity, we complete the proof in that case only.
We believe the generalization should work, because the proof for the
case of $ U_q(sl_2)$ essentially amounts to factorization on a corner for
the 3D TQFT, which works for the other quantum groups as well.

In order to prove topological invariance, we need a set of moves which
relate any two triangulations of the same PL manifold. We are
fortunate to possess a very convenient set of moves, as a result of
recent work by Pachner, in which he showed that triangulated manifolds
are equivalent if and only if they are bispherically equivalent [13].
(We wish to thank W.B.R. Lickorish for bringing this to our attention.)

The translation of our formula from 3D TQFT into the recombination
diagrams of quantum SU(2), as defined in [12], requires a great deal
of delicacy. We have found that the best way to define an invariant
of 4 manifolds in terms of the expressions in [12] is to order the
vertices of the triangulation and to make a consistent choice of
splittings, where each tetrahedron has its odd and even faces joined
separately, then connected in the middle. (By odd (even) faces, we
mean the faces which result from removal of the first and third
(second and fourth) vertices of the tetrahedron under the ordering). It is then
necessary
to normalize the diagram with an appropriate product of quantum dimensions.

In translating the trivalent graph associated to the 4-simplex into
a recombination diagram, we make the choice that the odd half of each
tetrahedron always comes from above. Thus, we associate to each
ordered 4-simplex the following graph, or its reflection depending on
whether
the orientation induced by the ordering agrees with the orientation on
the
4-manifold.

\vspace*{3in}

\bigskip

\centerline{ Figure 3}
\smallskip

With this choice,Ooguri's 6J symbols become trivial, and the formula,
with the proper normalization, takes the following form.

\bigskip

\[ \sum \hspace{1ex}
N^{\# vertices - \# edges} \prod_{faces} dim_q(j)
\prod_{tetrahedra} dim_q^{-1}(p) \prod_{4-simplexes} 15J_q \hspace{.2in}(*) \]
\bigskip

\noindent where the sum ranges over all assignments of spins to the
faces and tetrahedra of the triangulation and $j$ represents the spin
labelling a face, $p$ represents the spin labelling the cut interior
to a tetrahedron, $dim_q$ is the quantum dimension [12], and $ N$ is
the sum of the squares of the quantum dimensions.

It is now a direct matter of computation to check that our expression
is invariant under Pachner's moves. Basically, the moves work because
each move consists in replacing one half of the boundary of a
5-simplex by the other half.

When computing a sum of products of diagrams, in which an object in
one diagram is identified with one in another diagram, we can use some
simple recombination rules, which were originally discovered in the
representation theory of SU(2), but hold equally well for quantum
SU(2)
(and have analogues for any MTC).
Graphically, these rules are as follows:

\vspace*{6in}

\centerline{Figure 4}
\smallskip

Notice that in the preceeding figure, we have
modified the convention of Kirillov and Reshetikhin, and use maxima
and minima to denote mutiples of the usual contraction of indices maps. Our
modified normalizations are related to those in [12] as follows

\vspace*{2in}

\centerline{Figure 5}
\smallskip

Pachner's moves consist of replacing one half of the boundary of a
5-simplex
with the other half.  Observe that each half of the boundary of a
5-simplex
is a 4-ball, and they share a common (triangulated) $S^3$. The
verification
of invariance under Pachner's moves consists of using the
diagrammatic recombination formulas of Figure 4 to show that the
contribution to $(*)$ of either half of the boundary of a 5-simplex
reduces
to the evaluation of the invariant of labelled surfaces with trinion
decompositions on the surface in the common $S^3$ obtained by gluing
4-holed
spheres in each tetrahedron.

The computations are straightforward. The only difficulty was finding
the correct generalized 15J symbol and normalizing factors so that the
reductions can be carried out.

Given this description, it is probably easier for the reader to
reproduce the computation than for us to write it down.
\bigskip

{\bf4. Implications and Extensions}

\smallskip

Even for a MTC with a very small number of irreducibles, the number
of terms in our formula is too great for paper and pencil
computation. In simple cases, the use of diagrammatic formalism will
allow for hand calculation. We shall endeavor to make some computer
calculations in
later work.
\smallskip

{\bf Mathematical Implications}
\smallskip

There is a standard technique for turning an invariant based on a
decomposition into tetrahedra into a TQFT [14]. Thus, we have an
invariant of 4 manifolds of the same algebraic structure as the
one conjectured for Donaldson Floer theory. We are led naturally
to two questions: what do our new invariants tell us about smooth
4-manifolds? and how closely are they related to Donaldson-Floer
theory?

Witten [8], gives a gauge fixed form of a formal lagrangian for
Donaldson-Floer theory,  which is supersymmetric.
We could attempt to imitate this by using the representations of
quantum supergroups to construct a supersymmetric MTC, which might
bring us closer to Donaldson-Floer theory.

It would be interesting to compute our invariants, for the MTCs at
small roots of unity, on the examples of 4-manifolds which are known
to be homeomorphic but not diffeomorphic [15].
This calculation would require nothing difficult, except extensive
computation.
\smallskip

{\bf Mathematical Extensions}
\smallskip

It is rather surprizing that a 4D TQFT can be constructed from a
MTC. The form of the invariant, with corrections associated to
simplices of different dimensions, is suggestive of a picture in
which one associates categorical structures of different levels
to simplices of different dimensions. One is tempted to think that
the MTC is functioning here as a 2-category with one object, or even
as a 3-category with 1 object and one morphism. This would explain the
most surprizing aspect of Ooguri's suggestion, namely that spins
appear on faces, rather than on edges, as in a gauge theory.

We are therefore led to conjecture that the invariant we calculate
here is a special case of one calculated from a 2-category.

It is interesting to wonder whether the new 2-categories related
to the canonical basis for quantum groups [16, 17] yield any
interesting data in 4D topology.

The technique of attaching surfaces to triangulations we use is
similar to the standard technique for producing a Heegaard splitting
of a 3-manifold from a triangulation [22]. A Heegaard splitting can
be thought of as a handlebody decomposition, in which the attaching of
the 2-handles to the 1-handles is the surface map. One has the feeling
that the formulation of our 4D invariant in terms of triangulations
could be more elegantly restated in terms of handlebodies.
\smallskip

{\bf Physical Implications}
\smallskip

The invariants of knotted graphs which are constructed from MTCs
are closely related to spin networks [18, 19]. Spin networks are
an old idea due to Penrose, which admit an interpretation as a
discretized form of 3D quantum gravity. Furthermore, the Chern-Simons
lagrangian, which Witten used to formally produce 3D TQFT
[20], is also a state for quantum general relativity in the
Ashtekar formalism [21]. It is a natural goal to try to relate
the spin network picture to a geometric interpretation of
Chern-Simons-Witten theory as a state for quantum general relativity.

The spin network picture allows us to rewrite the CSW invariant as a
sum over labelings of a triangulation of a region in a 3-manifold,
which can be interpreted as a sum over discretized 3-geometries.[18,
21].

One is faced with the thorny problem of reinterpreting the CSW
state as a 4D picture, ie, of reintroducing time into the picture.
It is very suggestive to note that the construction outlined in this
paper produces a 4D TQFT which is so closely related to a 3D one, and
which comes from labelling a triangulation of the 4-manifold, thus
suggesting a sum over discretized geometries of the 4-manifold.

In the best of possible worlds, Einstein's equation would appear in a
classical limit of a 4D summation, just as flat geometries appear in
[21].

\newpage

{\bf REFERENCES}
\smallskip

[1] V. Jones, {\em A Polynomial Invariant Of knots via Von Neumann Algebras}
Bull. Am. Math. Soc. {\bf 12} (1985) 103-111.

\smallskip

[2] N. Reshetikhin and V. Turaev, {\em Ribbon Graphs and Their Invariants
Derived From Quantum Groups}, Comm. Math. Phys. {\bf 127} (1990) 1-26.

\smallskip

[3] L. Crane, {\em 2-d Physics and 3-d Topology}, Commun. Math. Phys.
{\bf 135} (1991) 615-640.

\smallskip

[4] L. Crane, {\em Quantum Symmetry, Link Invariants and Quantum Geometry},
in Proceedings XX International Conference on Differential Geometric
Methods in Theoretical Physics, Baruch College (1991)

\smallskip

[5] G. Moore and N. Seiberg, {\em Classical and Quantum Conformal Field
Theory}, Commun Math. Phys. {\bf 123} (1989) 177-254.

\smallskip

[6] P. Freyd and D. Yetter, {\em Braided Compact Closed Categories
with Applications to Low-Dimensional Topology}, Adv. in Math. {\bf 77}
(1989) 156-182.

\smallskip

[7] M. Atiyah, {\em New Invariants For Three And Four Manifolds} in The
Mathematical Heritage Of Hermann Weyl, AMS (1988).

\smallskip

[8] E. Witten, {\em Topological Quantum Field Theory}, Comm. Math. Phys.
{\bf 117} (1988) 353-386.

\smallskip

[9] C. Taubes, {\em $L^2$ Moduli Spaces On 4-Manifolds With
Cylindrical Ends, I}, Harvard preprint.

\smallskip

[10] H. Ooguri, {\em Topological Lattice Models in Four Dimensions},
Kyoto University preprint.

\smallskip

[11] V. Turaev and O. Viro, {\em State Sum Invariants of 3- Manifolds and
Quantum 6-J Symbols}, Topology {\bf 31} (1992) 865-902.

\smallskip

[12] A. Kirillov and N. Reshetikhin, {\em Representations Of The Algebra
Uq(sl(2)), q-Orthogonal Polynomials And Invariants of Links} in
Infinite Dimensional Lie Algebras and Groups, World Scientific (1989).

\smallskip

[13] U. Pachner, {\em P.L. Homeomorphic Manifolds Are Equivalent by
Elementary Shelling}, Europ. J. Combinatorics {\bf 12} (1991) 129-145.

\smallskip

[14] D. Yetter, {\em Topological Quantum Field Theories Associated To Finite
Groups and Crossed G-Sets}, Journal of Knot Theory and its
Ramifications {\bf 1} 1 (1992) 1-20.

D. Yetter, {\em Triangulations and TQFT's}, to appear.

\smallskip

[15] S.Donaldson, {\em Irrationality and The h-Cobordism Conjecture},
J. Diff. Geo. {\bf 26} (1987) 141-168.

\smallskip

[16] L. Crane and I. Frenkel, {\em Hopf Categories and Their
Representations}, to appear.

L. Crane and I. Frenkel {\em Categorification and the Construction of
Topological Quantum Field Theory},  to appear in Conference Proceedings
AMS Conference on Geometry Symmetry and Physics, Amherst, MA, Summer 1992.

\smallskip

[17] D. Kazhdan and Y. Soibelman, {\em Representations of The Quantized
Function Algebras, 2-Categories and Zamolodchikov Tetrahedra Equation},
Harvard preprint.

\smallskip

[18] L. Crane, {\em Conformal Field Theory, Spin Geometry, and Quantum
Gravity}, Phys.  Lett. B {\bf 259} 3 (1991).

\smallskip

[19] R. Penrose, {\em Angular Momentum; an Approach to Combinatorial Space
Time}, in Quantum Theory and Beyond, ed T.Bastin, Cambridge.

\smallskip

[20] E. Witten, {\em Quantum Field Theory and The Jones Polynomial},
Comm. Math. Phys. {\bf 121} (1989) 351-399.

\smallskip

[21] G. Ponzano and T. Regge, {\em Semiclassical Limits of Racah Coefficients},
in Spectroscopic and Group Theoretical Methods in Physics, ed. F.
Bloch, North-Holland.

\smallskip

[22] D. Rolfson, {\em Knots and Links}, Publish or Perish, (1976).

\end{document}